\documentclass[prl,aps,twocolumn,superscriptaddress,showpacs]{revtex4}
\usepackage{amsmath}
\usepackage{graphicx}
\usepackage{bm}

\newcommand{\rl}{\rangle\!\langle}
\newcommand{\hk}{\bm{k}}

\newcommand{\bkl}{b_{\bm{k},\lambda}}
\newcommand{\bkld}{b_{\bm{k},\lambda}^{\dag}}

\newcommand{\bmkl}{b_{-\bm{k},\lambda}^{\dagger}}

\newcommand{\sumkl}{\sum_{\bm{k},\lambda}}
\newcommand{\wk}{\omega_{\bm{k}}}
\newcommand{\wkl}{\omega_{\bm{k},\lambda}}

\begin{document}

\author{{\L}ukasz Marcinowski}
\affiliation{Institute of Physics, Wroc{\l}aw University of
Technology, 50-370 Wroc{\l}aw, Poland}
\author{Katarzyna Roszak}
\affiliation{Institute of Physics, Wroc{\l}aw University of
Technology, 50-370 Wroc{\l}aw, Poland}
\affiliation{Department of Condensed Matter Physics,
Faculty of Mathematics and Physics, Charles University,
12116 Prague, Czech Republic}
\author{Pawe{\l} Machnikowski}
\email{pawel.machnikowski@pwr.wroc.pl}
\affiliation{Institute of Physics, Wroc{\l}aw University of
Technology, 50-370 Wroc{\l}aw, Poland}

\title{Singlet-triplet dephasing in asymmetric quantum dot molecules}

\begin{abstract}
We discuss pure dephasing of singlet-triplet superpositions in
two-electron double quantum dots due to elastic phonon scattering. We
generalize our previous results to a system built of two non-identical dots. We
show that the asymmetry must be very strong in order to considerably
affect the dephasing rate.
\end{abstract}

\pacs{73.21.La, 03.65.Yz, 72.10.Di, 03.67.Lx}

\maketitle

Dephasing of singlet-triplet superpositions in two-electron quantum
dot molecules is important for a possible implementation of quantum
information processing in semiconductor systems \cite{loss98}. In a
recent work \cite{roszak09},
we showed that elastic phonon scattering via virtual transitions
to doubly occupied states, which is only possible in a singlet
configuration, induces distinguishability of spin configurations and,
therefore, leads to pure dephasing of spin superpositions. 
In Ref.~\cite{roszak09}, we studied a system of two identical
dots. However, as quantum dots
are artificial systems, one has to take into account
unavoidable inhomogeneity of dot parameters when modeling the
properties of the system. Therefore, in the present contribution, we
generalize our previous result and study the phonon-induced dephasing
process in an asymmetric QDM. We show that in the
asymmetric QDM, an additional dephasing channel appears, as compared to
the symmetric one. Nonetheless, the dephasing rate is very weakly
affected by the asymmetry unless the latter becomes very strong.

The system under consideration is composed of
two electrons in an asymmetric quantum dot molecule (QDM) built from two
different gate-defined quantum dots
\cite{petta05,koppens05}. The electrons are coupled to phonons by the
usual charge-phonon interactions (deformation potential and
piezoelectric couplings). We do not take any spin-environment
interactions into account (neither direct, with nuclear spins, nor
indirect, via spin-orbit coupling). The Hamiltonian of the system is then
\begin{equation*}
H=H_{\mathrm{DQD}}+H_{\mathrm{ph}}+H_{\mathrm{int}}.
\end{equation*}
The first term describes the electrons and has the form
\begin{eqnarray}
H_{\mathrm{DQD}}
&=&\Delta \epsilon\sum_{s}\left(  
a_{\mathrm{L}s}^{\dag}a_{\mathrm{L}s}-a_{\mathrm{R}s}^{\dag}a_{\mathrm{R}s}
\right)\nonumber\\
&&-t_{1}\sum_{s}\left(  a_{\mathrm{L}s}^{\dag}a_{\mathrm{R}s}
+\mathrm{h.c.} \right) \nonumber \\
&&+\frac{1}{2}\sum_{s,s'}\sum_{i,j,k,l}
V_{ijkl}a_{is}^{\dag}a_{js'}^{\dag}a_{ks'}a_{ls},
\label{H-DQD}
\end{eqnarray}
where $a_{is},a_{is}^{\dag}$ are the electron annihilation and
creation operators with $i=\mathrm{L,R}$ denoting the left and right
dot, respectively, and $s=\,\uparrow,\downarrow$ labeling the spin
orientation. The first term in Eq.~(\ref{H-DQD}) accounts for the
energy difference between single-electron states in the two dots. The
second term represents single-particle inter-dot tunneling.  The third term
describes the Coulomb interaction, with $V_{ijkl}=V_{jilk}=V_{klij}=V_{lkji}$
(the wave functions may be chosen such that the matrix elements are
real). 

The Hamiltonian of the phonon reservoir is given by
$H_{\mathrm{ph}}=\sum_{\bm{k},\lambda}
\hbar\wkl\bkld\bkl,$
where $\bkl,\bkld$ are annihilation and creation operators for a
phonon from a branch $\lambda$ with a wave vector $\bm{k}$ and
$\hbar\wkl$ are the corresponding energies. 
The electron-phonon interaction is described by
\begin{displaymath}
H_{\mathrm{int}}=\sum_{s,i}\sumkl
F_{i}^{(\lambda)}(\hk)a_{is}^{\dag}a_{is}(\bkl+\bmkl),
\end{displaymath}
where 
$F_{\mathrm{L/R}}^{(\lambda)}(\hk)$ are coupling
constants. We use the usual coupling
constants for confined charges (see Refs.~\cite{grodecka08a,roszak09}
for explicit expressions).

The Hamiltonian $H_{\mathrm{DQD}}$ can be diagonalized in a simple
way. The resulting singlet eigenstates can be written in the form
\begin{eqnarray*}
\lefteqn{|S_{0}\rangle  =}\\
&&\cos\frac{\phi}{2}\left[ 
\cos\frac{\theta}{2}|(1,1)S\rangle-\sin\frac{\theta}{2}|(+)S\rangle 
\right] -\sin\frac{\phi}{2}|(-)S\rangle\\
\lefteqn{|S_{1}\rangle =}\\
 &&-\sin\frac{\phi}{2}\left[ 
\cos\frac{\theta}{2}|(1,1)S\rangle-\sin\frac{\theta}{2}|(+)S\rangle 
\right] +\cos\frac{\phi}{2}|(-)S\rangle\\
\lefteqn{|S_{+}\rangle =  \sin\frac{\theta}{2}|(1,1)S\rangle
+\cos\frac{\theta}{2}|(+)S\rangle,} 
\end{eqnarray*}
where $|(\pm)S\rangle=[|(2,0)S\rangle\pm |(0,2)S\rangle]/\sqrt{2}$ and
$|(m,n)S\rangle$ denotes the configuration with $m$ electrons in the
left dot and $n$ electrons in the right one. 
The corresponding energies are denoted by $E_{0},E_{1},E_{+}$.

Here $\theta$ describes
mixing of the $|(1,1)S\rangle$ and $|(+)S\rangle$ configurations due
to tunneling and Coulomb interactions.
In the
present work we fix $\theta=0.5$, which corresponds to
$t_{1}/(E_{1}-E_{0})=0.2$. 

The angle $\phi$ accounts for state
mixing due to system asymmetry. The latter may be due to a
difference between the energies of doubly occupied configurations in
the two dots or to asymmetry of off-diagonal Coulomb
elements. Accordingly, we define two parameters
\begin{eqnarray*}
\Delta & = & 2\Delta\epsilon+\frac{V_{\mathrm{RRRR}}-V_{\mathrm{LLLL}}}{2},\\
\eta & = &
\frac{V_{\mathrm{RLRR}}+V_{\mathrm{LRRR}}-V_{\mathrm{LRLL}}-V_{\mathrm{RLLL}}}{2}.
\end{eqnarray*}
Neglecting the exchange terms compared to direct Coulomb energies, one
finds $\sin\phi\approx
[\sin(\theta/2)\Delta-\cos(\theta/2)\eta]/(E_{1}-E_{0})$.

The carrier-phonon coupling for the two-electron system is written in
the eigenbasis of $H_{\mathrm{DQD}}$,
\begin{displaymath}
H_{\mathrm{int}}=\sum_{\alpha,\beta,\bm{k},\lambda}
F_{\alpha\beta}^{(\lambda)}(\bm{k})|\alpha\rl \beta|
\left(\bkl+\bmkl \right),
\end{displaymath}
with $\alpha,\beta=S_{0},S_{1},S_{+}$.
Then, the phonon spectral densities involving the low-energy state
$|S_{0}\rangle$ are found,
\begin{displaymath}
R_{\alpha\beta}(\omega)=\sum_{\bm{k},\lambda}
F_{S_{0}\alpha}^{(\lambda)}(\bm{k})F_{S_{0}\beta}^{(\lambda)*}(\bm{k})
|n(\omega)+1|\delta(|\omega|-\wk),
\end{displaymath}
where $\alpha,\beta=S_{1},S_{+}$ and
$n(\omega)$ is the Bose distribution.
In contrast to thy previously studied symmetric case, two
channels of phonon scattering are now present, corresponding to the
two different excited states ($S_{1}$ and $S_{+}$) through which
the phonons can scatter. The interference of different scattering
paths is reflected by the presence of the
whole family of spectral densities out of which only one,
$R_{S_{1}S_{1}}(\omega)$ survives for $\phi\to 0$ and reduces to the spectral
density found in Ref.~\cite{roszak09}.

Following the method worked out in our previous paper \cite{roszak09},
we calculate the dephasing rate using the 4th order
time-convolutionless equation for the
evolution of the density matrix describing the state of the
two-electron system. From this, we extract the Markov limit, which
yields the dephasing rate due to two-phonon
(scattering) processes. Similarly as in the symmetric case
\cite{roszak09}, this process dominates over the one-phonon real transitions
at low temperatures. The resulting dephasing rate is given by
\begin{eqnarray*}
\gamma & = &
\pi\mathcal{P}\sum_{\alpha,\beta}\int_{-\infty}^{\infty}d\omega 
\left[
\frac{R_{\alpha\beta}(\omega)R_{\alpha\beta}(-\omega)}{
(\omega+\omega_{\alpha})(\omega_{\beta}-\omega)}\right.\\
&&\left.+\frac{R_{\alpha\beta}(\omega)R_{\alpha\beta}(-\omega)
-R_{\alpha\beta}(\omega_{\alpha})R_{\alpha\beta}(-\omega_{\beta})}{
(\omega-\omega_{\alpha})(\omega-\omega_{\beta})}\right],
\end{eqnarray*}
where $\omega_{\alpha}=(E_{\alpha}-E_{0})/\hbar$ and $\mathcal{P}$
denotes the Cauchy principal value.
The above equation generalizes the previously found formula to the
asymmetric case. 

In the calculations, we set the energy differences
$E_{1}-E_{0}=0.9$~meV, $E_{+}-E_{0}=1.0$~meV, the inter-dot distance
$D=300$~nm, and
use the material parameters as in Ref.~\cite{roszak09}.

\begin{figure}[tb]
\begin{center}
\unitlength 1mm
\begin{picture}(85,55)(0,5)
\put(0,0){\resizebox{85mm}{!}{\includegraphics{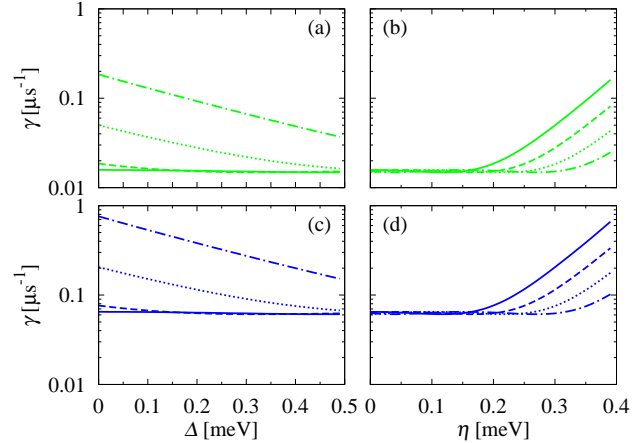}}}
\end{picture}
\end{center}
\caption{\label{fig:result}
Dephasing rate: (a,c) as a function of $\Delta$ for $\eta=0$ (solid
line), 0.2~meV (dashed line), 0.3~meV (dotted line), and 0.4~meV
(dash-dotted line); (b,d) as a function of $\eta$ for $\Delta=0$ (solid
line), 0.2~meV (dashed line), 0.4~meV (dotted line), and 0.6~meV
(dash-dotted line). In (a) and (b), $T=0.5$~K; in (c) and (d), $T=1$~K.}
\end{figure}

The pure dephasing rates resulting from the scattering
process are shown in Figs.~\ref{fig:result}(a,c) as a function of
$\Delta$ for a few values of $\eta$ and in
Fig.~\ref{fig:result}(b,d) as a function of $\eta$ for a few values of $\Delta$. 
In all the cases, the dephasing time varies from about 1 $\mu$s
to several $\mu$s. As long as the asymmetry is small, the dephasing rate
is rather insensitive to it. Only for strong asymmetry, the dephasing
rate grows considerably. One should keep in mind that the largest
values of $\eta$ shown in Fig.~\ref{fig:result} are rather
unrealistic since this parameter describes the variation of
exchange-like Coulomb terms which are themselves small. In a real
structure, $\eta$ is likely to remain within a tenth of meV. It is
clear from Fig.~\ref{fig:result} that in this parameter range the
dephasing rate is almost completely insensitive to $\eta$, as well as
to $\Delta$.

This work was supported in part by the Czech Science Foundation (Grant
no. 202/07/J051).

%\bibliographystyle{prsty}
%\bibliography{abbr,quantum}

\end{document}